\documentclass[12pt]{article}
 \usepackage{paralist} % item spacing \begin{compactitem} ...
     \usepackage{placeins} %keeps floats(tables,figures) `in their place'
     \usepackage{amsmath,amsthm,amssymb}
     \usepackage{epsfig}
      \usepackage{wrapfig} % wrap text around figures
      \usepackage{sidecap} % caption side of table and figures
     \usepackage{psfrag}
     \usepackage{tabls}
     \usepackage{supertabular}
\usepackage[square, numbers, sort&compress]{natbib}
     \usepackage[top=2.5cm, bottom=2.5cm, left=2.5cm, right=2.5cm]{geometry}
\usepackage{indentfirst} % indent firts paragraph after section
\usepackage{colortbl} % If before fancyheadings there is a problem
\definecolor{niceyellow}{rgb}{0.98,0.92,0.73}
\usepackage{fancyhdr}
\usepackage{fancybox}
\usepackage{hyperref} 
    \newcommand{\myfecha}{\today}
\makeatletter
\newcommand\ackname{Acknowledgements}
\if@titlepage
  \newenvironment{acknowledgements}{%
      \titlepage
      \null\vfil
      \@beginparpenalty\@lowpenalty
      \begin{center}%
        \bfseries \ackname
        \@endparpenalty\@M
      \end{center}}%
     {\par\vfil\null\endtitlepage}
\else
  \newenvironment{acknowledgements}{%
      \if@twocolumn
        \section*{\abstractname}%
      \else
        \small
        \begin{center}%
          {\bfseries \ackname\vspace{-.5em}\vspace{\z@}}%
        \end{center}%
        \quotation
      \fi}
      {\if@twocolumn\else\endquotation\fi}
\fi
\makeatother
\begin{document}
%------------------------------------------------------------
\begin{center}
%\textbf{
%{\Large{Working Paper}} \\
%\medskip
%---------
\textbf{
{\Large{Enhancing Reasoning Skills in the Process of Teaching and
Learning of Physics via Dynamic
Problem Solving Strategies: a Preparation for Future Learning}}\\
\medskip
{\Large{
Sergio Rojas \\
Departamento de F\'{\i}sica, Universidad Sim\'on Bol\'{\i}var \\ Venezuela.
}}
}
\end{center}

\begin{abstract}
The large number of published articles 
in physics journals under the title  ``Comments on $\cdots$'' 
and ``Reply to $\cdots$'' is indicative that the conceptual
understanding of physical phenomena is very elusive and hard to grasp
even to experts, but it has not stopped the development of Physics. In fact, 
from the history of the development of Physics  one 
quickly becomes aware that, regardless of the state of conceptual
understanding, without quantitative reasoning Physics would
have not reached the state of development it has today. 
Correspondingly, quantitative reasoning and problem solving skills are
 a desirable outcomes
from the process of teaching and learning of physics.
Thus,
supported by results from published research,
we  will show
evidence that a well structured problem solving strategy taught as a dynamical
process offers a feasible way for students to learn physics quantitatively
and conceptually, while helping them to reach 
 the state of an \emph{Adaptive Expert} highly
skillful on innovation and efficiency, a desired outcome from the perspective
of a \emph{Preparation for Future Learning} approach
of the process of teaching and learning Physics effectively.
\end{abstract}
%------------------
\section{Introduction}

From the perspective
of a \emph{Preparation for Future Learning} approach
of the process of teaching and learning physics effectively,
a contemporary view
encourage
addressing the process so that students become
\emph{adaptive experts} \cite{SchwartzEtAll:2005,HatanoInagaki:1986},
who are individuals highly efficient in applying (transferring) what they know
to tackle new situations and are also extremely capable of innovation
in the sense of being able to inhibit inadequate
blocking
``off the top of the head processes''
or ``to break free of well-learned routines''  so that they can move to
new learning episodes by finding, perhaps ingenious,
 ways to approach first time situations.

In this regard, it is undeniable that
 \emph{Physics Education Research} (PER) and psychological research
on learning and instruction have
made available a good deal of teaching strategies
which are helpful in reaching the aforementioned goal
 (a few such strategies are 
listed in \cite{HendersonDancy:2009}
and some of them have been reviewed elsewhere 
\cite{DocktorAndMestre:xxxx,LitzingerEtAl:2011,AmbroseEtAl:2010}).
Nevertheless, some controversial debates
in relation to the effectiveness of some of these \emph{Research-Based
Instructional Strategies} (RBIS) can also be found
in the literature
\cite{Sobel:2009a,Lasry:2009a,Sobel:2009b,Lasry:2009b,Glazek:2008,Klein:2007,KleinDebate1:2007,KleinDebate2:2007,KleinDebate3:2007}.

Nobel Prize winner Professor Carl Wieman has also called for
cautiousness when measuring RBIS teaching outcomes as one could create
illusions about what students actually
learn\cite{Wieman:2007a}. As mentioned in an article about transfer:
``standard methods of investigating transfer have
tended to depend on success-or-failure measures of participants' behavior 
on transfer
tasks designed with very specific performance expectations on the 
part of the investigator. These methods have failed to identify the 
productive knowledge that
students often do bring to bear on the tasks given to them.''\cite{Wagner:2006}
In this regard, Professor Sobel in more emphatic
``Yes, in a special (possibly
grant{}-supported) program, with smaller groups, with highly motivated
instructors and students, with less content, students might do well,
but that's not the real
world.''\cite{Sobel:2009b}
Or as warned by Professors Reif and Allen ``Because nominal expertise does not
necessarily imply good performance, one must be cautious in interpreting
cognitive studies of novices and experts for which 'experts' have been chosen
on the basis of nominal criteria. Data about such experts must be interpreted
cautiously to avoid misleading conclusions about thought processes leading
to good performance.''\cite{ReifAllen:1992}

In this panorama, 
of particular concern considering the intrinsically quantitative 
nature of
physics is the fact that
in physics classes 
students should actually be trained to apply what they have learned
in their math classes 
and the aforementioned opinions might be a result of the fact that
many of the published papers on teaching and learning   
physics 
seem to overemphasize the
importance of teaching conceptual physical
aspects
\cite{HoellwarthMoelter:2011,MualemEylon:2007,SabellaRedish:2007,Walsh:2007,HoellwarthEtAll:2005},
and to deemphasize the significance of standard mathematical
reasoning\cite{Taber:2009}, which are crucial for understanding 
physical processes, and
which are not stressed, or even taught, because, rephrasing a passage
from a recent editorial,
they interfere with the students' emerging sense of physical 
insight. \cite{Klein:2007} A view which is further
stressed in a physics textbook instructor manual 
(\cite{Knight:2008}, page 1-9):
``the author believes that for students struggling to grasp many
new and difficult concepts, too high a level of mathematics detracts
from, rather than aids, the \emph{physics} we want them to learn.
There is ample time in upper division courses for a more formal and
rigorous treatment. It's counterproductive to burden students with
unfamiliar and frightening mathematical baggage during their first 
exposure to the subject.'' And these kind of opinions seems to be reflected
in  outcomes obtained from the application of the relatively recently
 developed CLASS survey, which measures student's
beliefs about physics and learning physics, showing a decrease of roughly
$15\%$ (out of 397) after instruction on student's beliefs about problem 
solving in
physics
and a decrease of $12\%$ (out of 41) after instruction
on student's beliefs about the connection
between physics and mathematics 
(see respectively tables I and V of \cite{AdamsEtAl:2006}. Both courses were
calculus-based Physics I. In relation
to table I (N=397), the authors of the study mention that
``These are typical results for a first semester course----regardless 
of whether 
it is a traditional lecture-based course or a course with interactive
engagement in which the instructor does not attend to student's
attitudes and beliefs about physics."

Moreover, controversial outcomes coming from some highly publicized 
RBIS 
\cite{Ates:2007,Coletta:2008,Ates:2008,KostSmithEtAl:2010}
hardly help physics instructors in
finding suitable advice about how to approach the
teaching of physics in the most efficient way and an answer to the
question of how much time should be spent on intuitive, conceptual
reasoning and how much time in developing quantitative reasoning.

In view of the aforementioned facts, the aim of this paper is to
begin a discussion on how
the process of teaching and learning
physics via  \emph{dynamic problem solving strategies} can help
to tackle
not only the conceptual but also the quantitative reasoning
deficiencies
persistently
reported in the literature regarding the performance of students
in introductory and upper-division physics courses
\cite{WallaceChasteen:2010,Thompson:2009,Meredith:2008,MarshallCarrejo:2008,BoudreauxEtal:2008,Shaffer:2005,ReifAllen:1992,McDermott:1992:p1,Weeren:1982,LarkinEtAl:1980}.
It is obvious that both conceptual and quantitative reasoning
are desired skills students should acquire and develop in our
physics courses in order
to foster in them their willingness to explore more complex scientific
or engineering problems with confidence: \emph{a preparation
for future learning}.

The rest of the paper is organized as follows.
Recalling events
from the history of physics,
in the next section, 
\emph{Conceptual versus quantitative understanding}, 
we argue that
in spite of conceptual gaps in some key physical ideas
the development of physics did not stop
because of the quantitative nature of physics. We also present
in this section an empirical result showing how a student changes his/her
wrong initial (conceptual) intuition 
by using 
 quantitative
analysis when solving  a  problem about electrical circuits.
The next section discusses some of the needs for teaching students
the use and application of structural
 problem solving strategies.
The following section, before presenting
the conclusions, 
develops the central theme in the article: 
promoting deep approaches to learning via \emph{Dynamic problem
solving strategies}. An illustrative example is also discussed
in this section.
%------------------------------------------
\section{Conceptual versus quantitative understanding}

From a practical point of view, the conceptual understanding of the
principles of physics is a difficult and in some cases a very elusive
task. This is confirmed by the large body of research dealing with
``Student ideas about $\cdots$'',
 ``Student understanding of $\cdots$'', ``Students misconceptions about
$\cdots$'', ``students' misunderstanding of $\cdots$'' and so
forth.

For instance, in a 
study \cite{HrepicEtAll:2007}
performed in a rather highly suitable and exceptionally favorable
teaching environment, it was found that students answered correctly
some conceptual questions
on the nature of sound propagation in the same proportion 
before and after
receiving instruction on the subject. In that study, 
in addition to active teaching instruction, students also
watched at their own pace video-lectures on the nature of 
sound propagation
by an experienced instructor, Professor Paul Hewitt.
After the
pre-instruction test, the students
  were told 
 that each one of the test questions will be 
answered in these video-lectures.   
Similar results on students performance have also been reported 
from a study about why
the seasons change. 
Students maintained their conceptual misconceptions about the subject
even after watching a video that clearly explained the phenomena
\cite{AmbroseEtAl:2010}.

Perhaps more dramatic is the repeatedly reported case in which
students respond to conceptual questions about the behavior of physical
quantities the same way as students did at the beginning of 
the 80's \cite{Shaffer:2005}. 

Should we be surprised
by these findings? Not, at all. As mentioned by Ambrose and collaborators 
``It is important to recognize that conceptual change occurs gradually
and may not be immediately visible. Thus, students may be moving in the
direction of more accurate knowledge even when it is not yet apparent
in their performance'' \cite{AmbroseEtAl:2010}.
In fact, this is also observed in well trained physicists.
For example, great debates about
the proper understanding of the concept of physics have taken place 
among brilliant physicists. To be specific, 
in this regard
one could refer to
the conceptual debates in statistical and quantum 
physics and the electromagnetic 
theory \cite{Biro:2011,Plotnitsky:2010,BragaEtAll:2010,BacciagaluppiValentini:2009,Lindley:2001,Caneva:1980,Whittaker:1910}; 
the controversies
between Lorentz and Einstein on the conceptual understanding and the
meaning of the principles
of special relativity \cite{Janssen:2002};
 and in other areas \cite{Lange:2011,Bunge:1966}. 
What is even more relevant to PER is the fact that
many of these controversies persist
still today among expert physicists who invest a great deal of their time
thinking about and working with these matters
\cite{Swendsen:2011,Hobson:2009,Kampen:2008,Baierlein:2006,Lalo:2001,JacksonReplyRoche:2000,RocheReplyJackson:2000,Jackson:1999}.
Additionally, the large number of ``Comment on $\cdots$'' and
 ``Reply to $\cdots$'' articles in physics journals are also a reminder 
of the difficulty
of understanding and applying the concepts of physics.

Nevertheless, in spite of the aforementioned controversies on the 
conceptual understanding
of physical principles, the development of physics has not stopped. One could
argue that the reason for it 
is rooted in the fact that
``nature is too subtle to be described from any single point of view.
To obtain an adequate description, you have to look at things from
several point of views, even though the different viewpoints are
incompatible and can not be viewed simultaneously.''\cite{Dyson:1991} 
And,
 as a matter of fact,
 physics is fortunately a combination of two basic compatible viewpoints
 ``physical reasoning''
and ``quantitative reasoning''. 

Correspondingly, 
in spite of debates on the nature
and significance of the concepts of physics,
the intrinsic quantitative nature of physics is what has
propelled the development of physics, 
 helped by the
sometimes questioned 
\emph{Scientific Method} \cite{Chalmers:1999,Giunta:2001}.
 While the scientific method 
helps us to organize and test systematically
every single hypothesis (enhancing our conceptual understanding)
\cite{Kipnis:2010,Bassow:1991}, 
quantitative reasoning helps us to be precise
in which body of knowledge (mathematical models of the physical world) 
needs to be further developed: 
those \emph{de accord} or that are consistent with
observations and experiments \cite{Chalmers:1999,Peierls:1991,Peierls:1979}. 

To make the point clearer, 
one could think about the kind of progress physics
 would have reached had not Kepler
struggled to fit the orbit of Mars to an elliptical one,
stopping because of his lack of understanding (provided by Newton
around 80 years later) of why the orbits of the planets were
following the laws he was uncovering. Or think
about the course of knowledge had Galileo given up his view 
of doing experiments and finding mathematical explanations
for them
in favor of the Catholic Inquisition's conceptual ideas about the universe.
%---
%or think about the importance of Leverrier's 
%prediction of Neptune via Newton's Universal Law of Gravitation;
%---
Or think about the current state of development in physics had Planck (because
of lacking the respective conceptual understanding) restrained himself
from introducing (in 1900) the Planck's constant to 
resolve the ultraviolet
catastrophe. Or think about Einstein not continuing the
development of his Theory of General
Relativity because of the lack of conceptual understanding
for not  keeping in his theory
 the cosmological constant leading to a static universe.

But, have expert  physicists today overcome the conceptual understanding 
undermining
those developments? The answer in no. 
To be explicit in one case, one could see how physicists are still
trying to understand quantum mechanics at its deepest conceptual level,
a problem which arose more than one hundred years ago,
at the beginning of the last century, with
the introduction of Planck's constant. Yet, in spite of the conceptual
shortcoming, the mathematical formulation of quantum mechanics and
its refinements have allowed 
physics, regardless of the conceptual gap,
 to progress to levels that in today's world at many
physics and engineering
research centers, 
 researchers are making conclusive observations
about the nano-scale world for unanimated matter.

Thus, in each one the aforementioned cases, and of the many others that
can be cited,
there is no doubt that it was the quantitative analysis
undertaken by the scientist
involved that raised further the value
of scientific knowledge, even when at the time it was hard to provide
satisfactory conceptual explanations of the phenomena (such explanations
came much later, after further developments of the mathematical understanding
of each phenomena).
Without them one would be
talking today about ``philosophical or scriptural proclamations'' rather 
than scientific ones. 
Consequently, each one of these facts speak about the necessity of 
having our students
of science and engineering 
to become properly acquainted with quantitative reasoning in their early
training, even if they
are lacking deep conceptual understanding. In this way they will be able
to further deepen their understanding as they arrive to study upper
division courses.

%-----------
\begin{figure*}[htb]
\begin{center}
\leavevmode
  \includegraphics[height=.2\textheight]{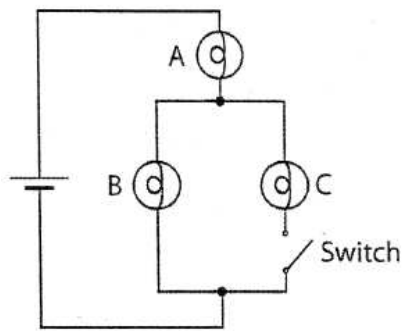}
\end{center}
\caption{The circuit contains an ideal battery, three identical light bulbs
and a switch. Initially the switch is open. After the switch closes, does
the brightness of bulb A increase? Explain.
This question was given
as a post-test written examination
to a total of 23 students at the University of Maine this year
(2011). Only two students
attempted to answer the question quantitatively. The others
applied rather unsuccessfully a \emph{Case Based Reasoning} 
approach \cite{Kolodner:1992}, trying to answer the question recalling 
from memory some classification patterns devised for conceptual 
understanding    
of this type of problems \cite{McDermott:1992:p2}. 
}
\label{fig:circuit}
\label{page:Figcircuit}
\end{figure*}
%-----------

In terms of the teaching and learning of physics,
 an empirical example of
how quantitative reasoning can help students to accurately 
reason conceptually, even though the students' initial intuition
  might be  wrong, can be seen in data
from a post-test written examination given to students
enrolled in a first year introductory physics course at the University
of Maine (UMaine). The formulation of the problem is shown in
Figure \ref{fig:circuit}. At the start,
  answering the question intuitively, one student
wrote that the brightness of bulb A should decrease.
Then the student went on to explain why it would happen that way as follows:
``The brightness should decrease
because the brightness of the bulb depends on the current of the circuit.
So when the switch is open to find the current of the two light bulbs in series
you would use the formula
$I_{\text{total}} = V_{\text{total}}/R_{\text{total}}$. So let say $V=12 v$
and each bulb acts as a resistor with $2\Omega$ of impedance. When it's
open there are two resistors in the circuits so $I = 12/(2+2)=3 A$
compared to when it's closed we simplify it to a series circuits so
the $R_{\text{total}}$ of the parallel would be $1 \Omega$ so
$1 \Omega + 2 \Omega = 3 \Omega$ as $R_{\text{total}}$.  $V=12 v$
so $12/3 = 4 A$ so when the switch is thrown there's more current and the
bulb is brighter''.

Analysis of this and other students' answers of this study will
be published elsewhere \cite{Hawkins:2011}.
Here
we should only notice
how the use of quantitative reasoning helped the student to
correct his/her initial (wrong) intuition
to the correct result that
after closing the switch bulb A becomes brighter.
Thus, this example provides
 evidence that the learning
with
emphasis on equations and stressing
the conceptual physical meaning of the respective
symbols in the equation is a very feasible task. In particular, 
proper guidance and additional training in applying physical equations
with understanding
will
help this student to reason analytically, using symbols,  instead of
 resorting to numerical
values (though we are not against this practice, specially in more
difficult situations), which is more useful
when dealing with situations on which physical intuition might 
fail \cite{Singh:2002} (assuming that resistance of bulb $A$ is 
known, a non-intuitive question
regarding the circuit of figure \ref{page:Figcircuit}
would be to ask about values for the resistance of bulbs $B$ and $C$ 
in
order to have a maximum power in that section of the circuit
when closing the switch. This is a kind of
open-ended problem which 
has multiple solutions.)

From the cognitive
point of view, the helpfulness of quantitative reasoning in conceptual
understanding can be grounded in the assumption
that
``...the use of mathematics in physics presupposes measurements. Measurements 
transform difficult to relate perceptual quantities into a common numerical 
ontology that supports precise comparisons and 
relations.''\cite{SchwartzEtAll:MathPhys:2005} 

Thus, while ``guiding students through a process of conceptual change
is likely to take time, patience, and creativity''\cite{AmbroseEtAl:2010},
students needs to becomes acquainted in applying what they have been
 learning quantitatively, so they can readily use that knowledge 
appropriately in
their more advanced courses. To make this possibility a reality
one needs to devote time to approaching the process of teaching and
learning physics with emphasis on equations and the meaning, not only
of the equation but also of each one of the symbols in the equation
(for example, the meaning and consequences of
$y=y_0 + m x$ 
when $m$ represents a constant acceleration are different from the
situation on which it ($m$) might represents
a constant speed. In each case,
 the symbols, $y$, $y_0$ and $x$, might have different meanings too. For
instance, in the case of $m$ representing constant acceleration, $x$
could represent either position, then $y$ and $y_0$ should
represent the square of a velocity, or time, then $y$ and $y_0$ should
represent a speed).
In this way, one could minimize PER findings on students being 
proficient in manipulating formulas without understanding  
the meaning of the symbols in the equation \cite{Sherin:2001}.
Correspondingly, as pointed out
by Professor Hewitt \cite{Hewitt:2011}
``Isn't teaching emphasis on symbols
and their meanings in an introductory [physics] course a
worthwhile effort?''
%-------------------------
\section{On why a problem solving strategy is needed}

Another worrying outcome from the learning and instruction of physics
and science research literature 
is the observed lack of ability by students
to apply a structured reasoning
methodology that could, among other things,
 help them  to identify the nature of a problem as well as
 the principles and 
quantitative models 
to solve it.
An example can be found in a study \cite{Walsh:2007}
reporting that
out of 22 students solving a set of six physics problems, 9
``Analyzes the situation based on required variables. 
Proceeds by choosing formulas based on the variables in a trial and error
manner.'', 6 ``Proceeds by trying to use the
variables in a random way.'', 2 ``Proceeds by trying to 'fit' the
given variables to those
examples.'', and 5 ``Plans and carries out solution
in a systematic manner based
on that analysis.'' Similar observations can be drawn from the
analysis of interview excerpts
reported in other studies \cite{BoudreauxEtal:2008,Wagner:2006,Sherin:2001}.

This lack of a structural way of reasoning is also
observed in the responses to the circuit
question presented in 
the previous section 
(see Figure \ref{fig:circuit}). In relation to the same question,
another 
student reasoned as follows:
``No, Decrease. Adding more resistance drop the current. Using the \#'s I
used, power decreases $\therefore$ Brightness goes Down." Writing the
answer the student wrote 
$\text{Brightness} = P = IR^2$ (she/he wrote the following
numbers next to every circuit element
$V=6 v$ for the battery and $2\Omega$ at each bulb). The student
continued with $A+B = 4 \Omega$ and $A + (1/B + 1/C) = 3 \Omega$. Then she/he
wrote $V=I R$;  $V/R = I \Longrightarrow 6v/4\Omega = 1.5 A$; 
$V/R = I \Longrightarrow 6v/3\Omega = 2 A$; $P = I R^2 = 1.5\cdot(4)^2 =24W$
and $2\cdot(3)^2 =18W$. 

In addition,  several studies have shown how initial thought processes 
block students'
thinking
 preventing them from going beyond a circular way of 
reasoning \cite{Kahneman:2002,BilalicEtAl:2008}.
To mention an example, in analyzing the interview of a student solving a physics
problem it is reported that ``Dee-Dee was very unwilling to give up her 
qualitative ideas about force and motion, even though she has already 
written down the correct algebraic form of Newton's second law. Her 
qualitative and quantitative dynamics knowledge appear to be associated 
(she articulated them very close together in time), but they have not 
been reconciled into a consistent knowledge 
structure.''\cite{SabellaRedish:2007}. 

Thus, in all of the mentioned cases we can observe
in most of the students 
the lack of a structured and systematic reasoning 
strategy which could
guide them to further analyze, verify,  and make sense of the solution
procedure they use when solving a problem. 
Additionally, this fact has also been reported in many studies comparing novice
and expert reasoning 
abilities \cite{Singh:2002,ReifAllen:1992,ReifHeller:1982,LarkinEtAl:1980}.
The most common observed behavior in students is the plugging of numbers
in equations without much hesitation.
As a matter of fact, in a study \cite{Ogilvie:2009} 
on which students were asked
to write down self-reflections on problem-solving one of them wrote
``Instead of learning the material and then doing the
assignments, I would just try to search for equations in the
text that would solve the problem, and if I couldn't figure it
out, just guess.'' Other student wrote 
``The way I approach physics problems is by looking for a
formula to follow. I will start a problem, look in my notes for
a formula, look on the discussion board for a formula, look
on the formula sheet for a formula, and lastly, look in the
book for a formula. This is probably exactly I will approach
some problems in Computer Science. If I need to implement
a function, I will need to find the syntax for the function and
a little excerpt saying what inputs it has and what outputs it
has.''

Unfortunately, that behavior is encouraged
in some instructional settings \cite{Hamed:2008} and
throughout the summary of equations at the end of each chapter on
commonly used textbooks.  And it is further stressed by the way
illustrative examples are worked out in those textbooks 
\cite{Hewitt:2011,Kamal:2011,Rojas:2010rmf}. 
Correspondingly, the 
problem solving methodologies
commonly found in many textbooks 
have been heavily criticized in PER 
literature \cite{KimPak:2002,DocktorAndMestre:xxxx}. And we agree with
 such criticism because what is preached and mostly illustrated in
commonly recommended textbooks for introductory physics
is a mechanical and static problem solving methodology: 
find an equation, plug in the numbers, and get the answer, 
taking away the joy of finding and exploring the different 
ways of solving a problem and the making sense not only of the solution
procedure but also of the obtained answer. 

More disappointing,  
textbooks 
examples end when the answer to the problem is found.
At most, in some few cases, there is a checking for dimensional consistency
of the result,
and in very rare cases one can find a discussion about the
feasibility of the obtained result.
But in general,
there is no further exploration of the solution procedure 
(i. e. whether or not it is a logically applicable to 
the situation at hand) and
no advice is given to students
on how they can be certain that the followed solution procedure is correct
(as a matter of fact, many wrong solution procedures found in textbooks 
have been
reported in the literature without being corrected
by textbook writers 
\cite{Saccomandi:2010,Bohren:2009,HutzlerEtAl:2004,Sandin:1973}).
More importantly,
 no
 advise is given to students regarding the fact
that some incorrect solution procedures
can lead to a correct solution \cite{Rojas:2010rmf}
or that some solution procedures
can be applied or transfered to solve other problems in different contexts
(just to mention one example, the computation of the gravitational and 
the electrical field of 
a mass and a charge distribution respectively
 share some similarities, but that is
not mentioned in most commonly used textbooks. Additional examples
on this matter are mentioned in \cite{Rojas:arxiv:2010}). 

Correspondingly,  students don't get trained
in developing a sense of solving problems by analogy. And they might 
even think that the mathematics used in physics is different from 
the mathematics
they learn in their calculus classes \cite{Rojas:2008}. Furthermore,
in most commonly used textbooks, physics
equations are not fully discussed to properly connect the physical or
conceptual meaning of each symbol with the place the symbol has
in the equation (i.e. why and what does it mean that the symbol is 
multiplying? or why and what does it mean that it appear as a negative
exponential factor? What happens if the value of a symbol increases?,
and so on). 
This lack of further analyzing the significance of the symbols in
equations might explain difficulty of students reasoning in
situations involving multivariable 
equations \cite{Thompson:2009,Kuhn:2009,Kuhn:2008,diSessa:2008}.
Moreover, the inadequacy of text-book worked-out examples to enhance
students learning has been made 
evident in a study by Chi and collaborators \cite{ChiEtAll:1989}. 
In the study it was found that the textbook worked-out examples did not
provide any clue for students make generalizations of the underlying
domain theory, a necessary training in order to help students
to develop important skills to
successfully transfer and apply the acquired knowledge in more complex
 contexts.  

Accordingly, the need for explicitly incorporating on the teaching
of physics a well structured \emph{dynamic problem solving strategy}
(introduced and explained in the next section)
 is thus justified.
As a matter of fact, in the same way as the \emph{Scientific Method} 
can be used to decide
among competing theories, a \emph{dynamic problem solving strategy} is a 
means by
which students can 
 guide their thoughts
 in deciding on the plausibility or reasonability
of each one of the steps taken when
 solving a (physics)
problem. Furthermore, a \emph{dynamic problem solving strategy} helps
students to make sense
of what they are learning by fitting it into what they already know or believe.
Via the questioning of intermediated
results and asking questions about what is being done, 
students can
detect flawed/wrong conceptual and/or computational procedures.
As students gradually make reflections on what is being done while solving
a problem, they
start to create or strengthen
 their own ``mental library'' of what works and what does not work.
As quoted by the Nobel Laureate in Economic Sciences (1978) 
Herbert Simon (cited in \cite{AmbroseEtAl:2010})
 ``\emph{Learning results from what the student
does and think and only from what the student
does and think. The teacher can advance learning only by influencing
what the student does to learn}.'' 

Thus, introducing high school and university students to a 
\emph{dynamic problem solving strategy} will enhance and strengthen 
their argumentative thinking skills
and will also help them to organize their reasoning skills by focusing 
their mental effort. It additionally could help to internalize early in 
students the fact
that (i) the process for solving scientific problems
requires creative
thought, the use of available resources (books, computers, articles, etc.), 
and personal interaction with peers and colleagues. (ii) There
may be no simple answer to questions that have been posed. In some
cases the outcome of a calculation can be contrary to what is expected
by physical intuition. In other instances an approximation that
 appears feasible turns out to be unjustified, or one that looks 
unreasonable turns out to be adequate.
(iii) There might be several competing ``correct'' answers based
on available knowledge and the careful and judicious  application 
of a \emph{dynamic problem solving strategy} is necessary to pick 
the correct one. 
And (iv) the full understanding 
of a problem and its solution requires both the quantitative formulation 
of the problem and the conceptual
meaning of the symbols that appear in the equation(s) describing 
the problem quantitatively. 
In short, the learning of a  \emph{dynamic problem solving strategy} would
be the starting point to generate a culture of reasoning and of making sense 
in the classroom, a necessary condition in
``preparation for future 
learning''
\cite{Osborne:2010,Fortus:2009,SchwartzEtAll:2005,HatanoInagaki:1986}.
%----------------------------
\section{A dynamic problem solving strategy}
An important issue to be resolved in PER is the lack of an integrative
theoretical framework that can make sense of the richness
of the large amount of empirical results 
collected via the many
\emph{Research Based Instructional Strategies}
(RBIS) generated in the field (for a partial listing of RBIS see 
\cite{HendersonDancy:2009}). In this regard, some obstacles
need to be overcome. In addition to controversial 
outcomes
mentioned in the introduction concerning some RBIS, 
important ongoing debates alluding to fundamental issues associated with
the psychology of learning physics
\cite{Slotta:2011,HammerEtAl:2011}
indicate that we are still far from reaching such a theoretical
framework. Until then, inspired by the scientific method framework,
we find it plausible to enhance the teaching and learning
of physics via \emph{dynamic problem solving strategies} in order to
 strengthen students' quantitative problem solving skills,
a necessity which is further stressed by the occurrence of
some disasters which have been associated to computing 
mistakes \cite{Feldman:2010,StrogatzEtAl:2005}.

In general terms, a \emph{dynamic problem solving strategy},
not to be confused with \emph{modes of reasoning} (see below), 
could be constructed from the following steps \cite{Rojas:2010rmf}:
(1) understand and describe the problem;
(2) provide a qualitative description of the problem;
(3) plan a solution;
(4) carry out the plan;
(5) verify the internal consistency and coherence of the equations used 
    and  the applied procedures; and 
(6) check and evaluate the obtained solution.

One step more or one step less, the aforementioned
\emph{problem solving strategy}  looks similar, you
might rightly wonder, to any other problem
solving strategy commonly found in textbooks and which have
been heavily criticized
in PER literature \cite{KimPak:2002}. In fact, as mentioned earlier 
we agree with those
criticism. For one additional reason,
once introduced, the strategy is not
consistently applied in the textbook illustrative examples, much
less in the student and instructor companion manuals. For another,
the steps of the problem solving strategy are presented as rigid steps
which need to be followed in the particular order they are written
and applied without connection or interaction between them
(certainly such an inflexible problem solving strategy can only
be of limited value). And finally,
textbooks end of chapter problems are
constructed and organized in such a way that students only need to find the
right equation to  plug-and-chug some numbers to get
the answer to the problem. Correspondingly, consequences
 of such strategies are reflected
in the findings of some PER studies (\cite{KimPak:2002} and 
references there in).

On the other hand, a study comparing a typical textbook problem 
solving strategy
with a consistently and coherently applied
 strategy reports on the overwhelming advantage of the latter
in relation to the former
\cite{HellerReif:1984}. 
To be specific, the authors of the comparative study propose
a prescriptive
theoretical model of effective human problem solving 
according to which the problem solving process embraces 
three major stages:
1) generation of an initial problem description (including qualitative
analysis) which is helpful in the construction of a problem solution;
2) obtaining the solution using appropriated methods; and 3) evaluation
and improvement of the solution. General comments on the nature
and significance of each stage 
are given in \cite{ReifHeller:1982}. 

In the referred work \cite{HellerReif:1984}, Professors Heller and Reif 
discussed further details of the controlled study they performed
comparing the quantitative problem solving performance of three groups:
one guided by their proposed problem solving strategy, the other
group guided by typical textbooks problem solving directions, and
a comparison group working without any external guidance.
In addition to showing the remarkable superiority of the
participants 
trained according to the
authors methodology,
three major lessons can be drawn from the study:
 a) that participants guided by a problem solving strategy
performed better than those who did not have any guidance; b) that
``completeness and explicitness of procedures for constructing initial
problem description'', missing from textbooks problem solving 
strategies, are crucial for attaining better problem solving performance;  
and c) that the participants implemented any systematic problem 
solving procedure
fairly easily once they became familiar with it, but to get to that level,
the procedure needs to be applied constantly,  
consistently and coherently because ``human subjects tend to be fallible
and distractable, prone to forget steps in a procedure or to disregard
available information.'' The authors comment that
after the resistance was overcome, 
some participants remarked that the steps ``really
work'' and that the problems seemed suddenly ``easy'' to solve,
showing in turn that students beliefs could be changed via a well
designed and applied problem solving strategy, while
at the same time strengthening their computational skills
(similar results are reported in a study in mathematics \cite{Higgins:1997}.
The author mentions that ``Compared with the students who 
had received traditional mathematics instruction, the students who had 
received problem-solving instruction displayed greater perseverance in 
solving problems, more positive attitudes about the usefulness of 
mathematics, and more sophisticated definitions of 
mathematical understanding.'') 

At this point we now turn to the topic of this section trying to answer
the question: 
what, then, makes a \emph{dynamic problem solving strategy}? 

First,
the implementation
of the methodology is not an inflexible static process (i.e. there 
is not a specific order
for implementing the considered steps and one can avoid some of the
steps or even add a new one). For instance, 
working via inductive reasoning one could start from the answer 
to an unknown problem
(i.e. an observed natural phenomena) and work backward to provide
a well posed problem whose solution is the observed answer. Students
can be trained in this process via well designed and worked out exercises by
 presenting them with
a careful and detailed
backward
analysis of the followed 
solution process. More importantly, verification of a solution procedure
via backward analysis,
in addition to
helping students
build confidence
in the obtained solution, also guides them to organize information in their
memory by making connections with
prior knowledge, a condition which increases the accessibility of useful
knowledge.

Thus, going back and forth in the application of a
\emph{dynamic problem solving strategy}
provides the required
feedback that experts apply when dealing with the solution
of new situations.
This flexibility helps
to avoid getting trapped or stuck by top-of-the-head thoughts, 
preformed convictions, or intuitions that don't help in going 
forward \cite{Kahneman:2002,BilalicEtAl:2008}. 
One just continue
developing and applying (mentally and/or in writing)
thought processes
until a solution is found, further analyzed and reconciled with
intuitions. We can either change the wrong intuition (i.e. as
one student did when solving the exercise
 of figure \ref{fig:circuit}) or strengthen
the correct one. Additional guidelines on what sort of analysis should
be included can be extracted from the work of Professors 
Polya \cite{Polya:1945}, Schoenfeld \cite{Schoenfeld:1980,Schoenfeld:1992}, 
and Reif \cite{Reif:2008}.
This way of building understanding can be contrasted with the mechanical
way in which textbook's illustrative examples are 
worked out.

Second,
proper application
of a 
\emph{dynamic problem solving strategy} 
requires the consistent use of
any applicable \emph{mode of reasoning}:
deductive and/or inductive reasoning; reasoning via analogy and/or via
counterexamples; reasoning by \emph{reductio ad absurdum}; 
 and many others, including rule, case, model, and
 collaborative-collective modes of reasoning. 
These modes of reasoning
are rarely mentioned in commonly used introductory physics textbooks. 
Nevertheless, in their mathematics courses
students might have already studied some of these modes of reasoning
(i.e. when proving by inductive reasoning the convergence of a sequence), 
and it is
in their physics courses where they should have the opportunity to further
explore such modes of reasoning
 from different perspectives, and became acquainted
with them.  

Third,
a very important 
skill in applying
a \emph{dynamic problem solving strategy} is the ability to ask questions.
Questioning, particularly at the higher cognitive
levels, is an essential aspect of problem 
solving. By asking questions while solving a problem
one becomes engaged in a process of
self-explaining components of the underlying theory being applied
to solve the problem and
that were not explicitly exposed when learning the theory. Asking
questions also helps in the
detection of ``comprehension failures'' and in taking action to overcome
them. As put by  Chi and collaborators ``Good students ask very
specific questions about what they don't understand. These specific
questions can potentially be resolved by engaging in self-explanations.''
 \cite{ChiEtAll:1989} 

Thus, at each step of a \emph{dynamic problem solving strategy}
 one needs to stop and ask oneself about the significance
of what has been done so far, trying to find meaningful associations between
the new knowledge being applied and related concepts one already might know.  
Examples of questions to be asked
constantly include:
how is this knew knowledge related to what I already know?;
in which context have I seen this problem before?;
in which context could I use this piece of knowledge?;
how are these seemingly disparate discrete pieces of knowledge 
functionally and causally related?; can the principles to be
applied be used in this situation?; is this approach the right one?; 
how can  I be certain of it?; are you sure you can do that?;
how could this procedure be wrong?.
Asking questions helps students to be fully confident that each
step given while solving a problem is a reasonable one and do not
involve contradictory or implausible assumptions. Also, asking questions helps
to seek
evidence to contradict an accepted fact and it helps to strengthen confidence
in a solution procedure if we can set out different scenarios to investigate 
the issues involved in a problem, as it requires engagement to identify
proper and reasonable ways to approach the problem and to think about
the interrelationships between the proposed approaches. 

Fortunately, the education research literature has provided a good deal
of research on how one can help students develop the
habit of asking questions
\cite{GraesserPerson:1994,ZeeMinstrell:1997,HarperEtAll:2003,Chin:2010a,Chin:2010b}. 
But, students will not
get the benefit of this process unless they
are explicitly taught how to use them. In this sense, being an intrinsic 
part of a 
\emph{dynamic problem solving strategy}, teaching it will develop in students
the habit of asking questions, a process
 through which students could
build new useful knowledge that
 they can then use in further developments as
they engage themselves in productive thinking and learning, not only
within the teaching and learning environment but also outside it.

We finalize this section in the hope of have made it clear that the
delivery of instruction following a \emph{dynamic problem solving
strategy} will create in students the habit of looking at problems
carefully and from multiple perspectives, choosing to be more mindful about
the making of sense of each solution procedure as they engage themselves
in the exploration of alternative ways of solving problems and of finding 
explanations. The development of such ways of thinking certainly requires time,
effort, practice, self-reflection, 
and feedback (from peers and the instructor). This is
how experts become experts, developing new intuition to cope properly 
with situations where there is no right answer.
%-------------------
\subsection{Illustrative example}

In general, most problem solving strategies found in the literature
only mention checking the feasibility of the obtained final
result. Since some wrong procedures could lead to right results,
we have found it necessary to include an additional step 
(step 5 mentioned at the beginning
of this section) in
constructing a \emph{dynamic problem solving strategy} so
faulty reasoning schemes could be detected \cite{Rojas:2010rmf}.

To illustrate the aforementioned thoughts, one could mention the 
fallacious application of the idea of 
\emph{separation of variables} (borrowed from solving partial differential 
equations \cite{Arfken:1985})
  to prove in two
dimensions the constant acceleration kinematic
relationships 
$v_x^2 = v_{0x}^2 + 2 a_x(x-x_0)$
and 
$v_y^2 = v_{0y}^2 + 2 a_y(y-y_0)$ (here
 $\vec{\mathbf{a}}= a_x \hat{\mathbf{x}} + a_y \hat{\mathbf{y}}$,
 $\vec{\mathbf{v}}= v_x \hat{\mathbf{x}} + v_y \hat{\mathbf{y}}$, and
 $\vec{\mathbf{r}}= x \hat{\mathbf{x}} + y \hat{\mathbf{y}}$
 represent respectively the acceleration, the velocity, and the displacement
of a particle, while
 $\vec{\mathbf{v}}_0= v_{0x} \hat{\mathbf{x}} + v_{0y} \hat{\mathbf{y}}$, and
 $\vec{\mathbf{r}}_0= x_0 \hat{\mathbf{x}} + y_0 \hat{\mathbf{y}}$ represents
respectively initial velocity and initial position of the particle. 
$\hat{\mathbf{x}}$ and $\hat{\mathbf{y}}$ are orthogonal unit vectors).

Starting from $d\vec{\mathbf{v}} = \vec{\mathbf{a}} dt$, dot product
both sides of this equation by $\vec{\mathbf{v}}$ and rearrange terms
in the form 
$d({v}^2/2) = \vec{\mathbf{a}}\mathbf{\cdot}\vec{\mathbf{v}}dt = 
\vec{\mathbf{a}}\mathbf{\cdot}d\vec{\mathbf{r}}$, which after integration,
considering $\vec{\mathbf{a}}=\text{constant}$, yields 
$v^2 = v_0^2 + 2\vec{\mathbf{a}}\mathbf{\cdot}(\vec{\mathbf{r}}
-\vec{\mathbf{r}}_0)$. This equation can be written in the form
\begin{equation}
v_x^2 - v_{0x}^2 - 2 a_x(x-x_0) = -(v_y^2 - v_{0y}^2 - 2 a_y(y-y_0)). 
\label{eq:separada}
\end{equation}

So far, there is nothing wrong in any of the given steps 
(in passing, let's mention that
by asking questions according to a \emph{dynamic problem solving strategy}
a student could gain conceptual understanding about the obtained 
relationship: when is it applicable?, why is time dependence not explicit?
can it be applied to free fall?, can it be applied to projectile motion?
if so, under which conditions?, etc.)

The faulty reasoning starts when a student, wrongly invoking the
separation of variables technique, considers that 
equation (\ref{eq:separada}) is in that form and, according to the 
technique, the student
proceeds to
write that $v_x^2 - v_{0x}^2 - 2 a_x(x-x_0) = \alpha$ and 
$v_y^2 - v_{0y}^2 - 2 a_y(y-y_0) = - \alpha$,
 with $\alpha=\text{constant}$.
Now, considering that at $\vec{\mathbf{r}} = \vec{\mathbf{r}}_0$, 
$\vec{\mathbf{v}} = \vec{\mathbf{v}}_0$, then $\alpha=0$ and
 the proof is obtained, the student believes. 

Thus, by means of following typical textbook
problem solving strategy, the student, without questioning any
of the solution steps, will consider that the proof is
correct because the 
obtained answer is correct, and to show that that it 
is so the student 
could point to any introductory
physics textbook on which the same result is obtained by other procedures. 

At this point one needs to mention that it is really hard to convince
students that despite leading to the correct answer, a solution procedure
could be mistakenly wrong. How, the student might wonder,
 can it be that a wrong
solution procedure could yield the right response?. That does not make
any sense!!! 

How can a \emph{dynamic problem solving strategy}
be helpful in making explicit the faulty reasoning? 
The trick is on step 5 mentioned at the beginning
of this section:   
once a solution procedure path has been established, one needs to
verify the consistency of each given 
step (a missing step
 from
any textbook problem solving strategy). And this can be attained
by working backward after the solution is obtained  and
by asking questions at each step, wondering and justifying the 
rightfulness of each one.

In this case, before calling
for the separation of variables technique, the student needs to ask about
whether the technique is applicable or not to the situation at hand. Since
the method of separation of variables
 comes from a differential equation context,
a first question to ask, following a \emph{dynamic problem solving strategy},
 is how to write equation (\ref{eq:separada}) so it looks like a differential
equation. This can be achieved by writing down the definition for each velocity 
component in the equation, namely
$v_x = dx/dt $ and $v_y = dy/dt$. After doing so it turns out
that the working equation is actually a nonlinear first order differential
equation on which the time variable $t$ is common to both sides of
the equation, making
it impossible to apply the separation of variables  procedure (which
requires that each side of the equation (\ref{eq:separada}) be a function
of only independent variables each to another. In this case both sides 
of the equation
turn out to be dependent of the time $t$ variable). 

Another way to settle the issue following a 
\emph{dynamic problem solving strategy} is to wonder whether
a counter example can be built to rule out the use of the separation
of variables in this case. It turns out that by using basic calculus
a very simple 
algebraic problem can be posed to see if the technique is or not applicable:
solve $(2x-x^2) = -(2y-y^2)$. 

Applying the proposed separation of variables,
instead of an infinite set of solutions, only a finite set of solutions
will be found  (namely 
$x=0$ and $y=0$ or $y = 2$; $x=2$ and $y=0$ or $y = 2$).  

Nevertheless, in spite of the evidence,
some students might still be reluctant to accept the wrongness
of their solution procedure and can start  to  
mention inexistent theoretical frameworks, like
 a \emph{separation of coordinates} without
being able to provide
any reference to backup such a reasoning.
In this situation, only self-reflection and careful analysis of
the quantitative procedure can be used as a way to solve the dilemma.

Episodes like this are not difficult to find in PER 
literature \cite{Sherin:2001},
and they can be described as ``reasoning to obtain the desired
result'', something that brings to memory what is
called Einstein's biggest mistake \cite{Weinberg:2005,Davies:1996},
which is associated  with the  
introduction by Einstein of a cosmic repulsion term 
in order to model, in spite of the mathematical evidence, a static universe.
Could this be called Einstein's conceptual misunderstanding? or is it
simply a case in which
 a good thought blocks a better one?\cite{Kahneman:2002,BilalicEtAl:2008}
%------------------------
\section{Concluding remarks}

  Understanding that physics is essentially a quantitative based
subject, 
in this article we have largely argue that the teaching 
of solving problems in introductory physics courses  
via a \emph{dynamic problem solving strategy}
can help students to develop quantitative reasoning
skills (necessary in the short term), while
 enhancing their capabilities to develop conceptual understanding
via the analysis of the equation(s) representing
physical phenomena followed by the correct interpretation of the
physical meaning of each symbol in the respective equation(s)
\cite{GaigherEtAl:2007} (for instance, as mentioned earlier, 
the meaning and consequences of
$y=y_0 + m x$
when $m$ represents a constant acceleration are different from the
situation on which it might represents
a constant speed. In each case,
 the other symbols, $y_0$ and $x$, might have different meanings too. For
instance, in the case of $m$ representing constant acceleration, $x$ 
could represent either position, then $y$ and $y_0$ should
represent the square of a velocity, or time, then $y$ and $y_0$ should
represent a speed).

At each step of a \emph{dynamic problem solving strategies}
students ask questions that guide them to be fully confident that each
step given while solving a problem is a reasonable one and do not
involve contradictory or implausible assumptions.
Thus, this view of enhancing the teaching and learning of introductory
physics courses via  \emph{dynamic problem solving strategies}
is de accord with the finding 
that ``Good students 
(those who have   greater  success at solving problems) tend to
study example-exercises in a text by
explaining and providing justifications for each action. 
That is, their explanations
refine and expand the conditions of an action, explicate the
consequences of an action, provide a goal for a set of actions, 
relate the consequences
of one action to another, and explain the meaning of a set of
quantitative expressions.''\cite{ChiEtAll:1989} 

Our proposal is further
 reinforced by the fact 
 that  teaching directly or indirectly (i.e.
via scaffolding tutoring)
 specific problem-solving strategies
improves students' scientific critical thinking and reasoning skills
\cite{Osborne:2010,MasonSingh:2010,Chi:2009,Nickerson:1988}.
As mentioned in a study 
 ``When students used the procedural specification, they did
so properly and obtained correct answers - although they did not always
implemented all steps explicitly and resorted to 
some shortcuts.''\cite{LabuddeReifQuinn:1988}
All of  that is in agreement with the fact
that 
learning to approach problems in a systematic way starts from 
learning the interrelationships among conceptual knowledge,
mathematical
skills and logical reasoning.\cite{HeronMeltzer:2005}
And
 it is in physics courses where
students can strengthen their quantitative reasoning 
skills and even become acquainted with innovative non-standard 
ways of solving problems
 \cite{OlnessAndScalise:2011,BernsteinFriedman:2009}.

Thus, since the process of teaching and learning can rarely be done
in a completely closed and controlled environment, a well learned
\emph{dynamic problem solving strategy}
will equip students with reasoning capabilities
to properly question advice (easily found  
on the
Internet and in 
some articles \cite{Taber:2009})
that encourages them to memorize results and to
ignore the mathematical analysis leading to those results.

In other words, 
considering that
``it is not possible to expect novice students to become more successful
problem solvers by simply telling them the principles which govern
the way experts sort physics problems." \cite{ChiEtAll:1989}
a well structured problem solving strategy  taught as 
a dynamical
process offers a feasible way for students to learn physics quantitatively
and conceptually, while helping them to reach 
 the state of an \emph{Adaptive Expert}, highly
skillful on innovation and efficiency: \emph{a preparation for future
learning}. 
%------------------
\begin{acknowledgements}
The author thanks the members of \emph{The Physics Education Research 
Laboratory} (PERLab) at UMaine, USA, where this work
was done while on Sabbatical in the Department
of Physics and Astronomy. 
\end{acknowledgements}
%----- References

%----- End References
\label{LastPage}


\begin{thebibliography}{118}
\expandafter\ifx\csname natexlab\endcsname\relax\def\natexlab#1{#1}\fi
\expandafter\ifx\csname url\endcsname\relax
  \def\url#1{{\tt #1}}\fi
\expandafter\ifx\csname urlprefix\endcsname\relax\def\urlprefix{URL }\fi

\bibitem[{Adams et~al.(2006)Adams, Perkins, Podolefsky, Dubson, Finkelstein, \&
  Wieman}]{AdamsEtAl:2006}
Adams, W.~K., Perkins, K.~K., Podolefsky, N.~S., Dubson, M., Finkelstein,
  N.~D., \& Wieman, C.~E. (2006).
\newblock New instrument for measuring student beliefs about physics and
  learning physics: The colorado learning attitudes about science survey.
\newblock {\em Phys. Rev. ST Phys. Educ. Res.\/}, {\em 2\/}, 010101:1--14.

\bibitem[{Ambrose et~al.(2010)Ambrose, Bridges, DiPietro, Lovett, \&
  K.}]{AmbroseEtAl:2010}
Ambrose, S.~A., Bridges, M.~W., DiPietro, M., Lovett, M.~C., \& K., N.~M.
  (2010).
\newblock {\em How learning works: Seven research-based principles for smart
  teaching\/}.
\newblock San Francisco, CA: Jossey-Bass.

\bibitem[{Arfken(1985)}]{Arfken:1985}
Arfken, G. (1985).
\newblock {\em Mathematical Methods for Physicist\/}.
\newblock Academic Press, 3rd. ed.

\bibitem[{Ates \& Cataloglu(2007)}]{Ates:2007}
Ates, S., \& Cataloglu, E. (2007).
\newblock The effects of students\char39{} reasoning abilities on conceptual
  understandings and problem-solving skills in introductory mechanics.
\newblock {\em Eur. J. Phys.\/}, {\em 28\/}, 1161--1171.

\bibitem[{Ates \& Cataloglu(2008)}]{Ates:2008}
Ates, S., \& Cataloglu, E. (2008).
\newblock Reply to \char39{}comment on ``the effects of students\char39{}
  reasoning abilities on conceptual understandings and problem-solving skills
  in introductory mechanics''char39{}.
\newblock {\em Eur. J. Phys.\/}, {\em 29\/}, L29--L31.

\bibitem[{Atkins(2007)}]{KleinDebate1:2007}
Atkins, L.~J. (2007).
\newblock Comment on ``{S}chool math books, nonsense, and the {N}ational
  {S}cience {F}oundation,'' by {D}avid {K}lein\cite{Klein:2007}.
\newblock {\em Am. J. Phys.\/}, {\em 75\/}, 773--775.

\bibitem[{Bacciagaluppi \& Valentini(2009)}]{BacciagaluppiValentini:2009}
Bacciagaluppi, G., \& Valentini, A. (2009).
\newblock {\em Quantum Theory at the Crossroads: Reconsidering the 1927 Solvay
  Conference\/}.
\newblock Cambridge University Press.
\newblock Available at \url{http://arxiv.org/abs/quant-ph/0609184v2}.

\bibitem[{Baierlein(2006)}]{Baierlein:2006}
Baierlein, R. (2006).
\newblock Two myths about special relativity.
\newblock {\em Am. J. Phys.\/}, {\em 74\/}(3), 193--195.

\bibitem[{Bassow(1991)}]{Bassow:1991}
Bassow, H. (1991).
\newblock Interdependency and the importance of errors in chemistry: How the
  search for a single error led to reexamination of the work of five nobel
  laureates and revised values for certain fundamental constants.
\newblock {\em J. Chem. Educ.\/}, {\em 68\/}(4), 273.

\bibitem[{Bernstein \& Friedman(2009)}]{BernsteinFriedman:2009}
Bernstein, M.~A., \& Friedman, W.~A. (2009).
\newblock {\em Thinking About Equations: A Practical Guide for Developing
  Mathematical Intuition in the Physical Sciences and Engineering\/}.
\newblock John Wiley \& Sons, New York.

\bibitem[{Bilalic et~al.(2008)Bilalic, McLeod, \& Gobet}]{BilalicEtAl:2008}
Bilalic, M., McLeod, P., \& Gobet, F. (2008).
\newblock Why good thoughts block better ones: The mechanism of the pernicious
  einstellung (set) effect.
\newblock {\em Cognition\/}, {\em 108\/}, 652 -- 661.

\bibitem[{Bir\'o(2011)}]{Biro:2011}
Bir\'o, T.~S. (2011).
\newblock {\em Is there a temperature?\/}.
\newblock Springer.

\bibitem[{Bohren(2009)}]{Bohren:2009}
Bohren, C.~F. (2009).
\newblock Physics textbook writing: Medieval, monastic mimicry.
\newblock {\em Am. J. Phys.\/}, {\em 77\/}, 101--103.

\bibitem[{Boudreaux et~al.(2008)Boudreaux, Shaffer, Heron, \&
  McDermott}]{BoudreauxEtal:2008}
Boudreaux, A., Shaffer, P.~S., Heron, P. R.~L., \& McDermott, L.~C. (2008).
\newblock Student understanding of control of variables: Deciding whether or
  not a variable influences the behavior of a system.
\newblock {\em Am. J. Phys.\/}, {\em 76\/}, 163--170.

\bibitem[{Braga et~al.(2010)Braga, Guerra, \& Reis}]{BragaEtAll:2010}
Braga, M., Guerra, A., \& Reis, J. (2010).
\newblock The role of historical-philosophical controversies in teaching
  sciences: The debate between biot and amp\`ere.
\newblock {\em Sci. \& Educ.\/}, (pp. 1--14).

\bibitem[{Bunge(1966)}]{Bunge:1966}
Bunge, M. (1966).
\newblock Mach's critique of newtonian mechanics.
\newblock {\em Am. J. Phys.\/}, {\em 34\/}(7), 585--596.

\bibitem[{Caneva(1980)}]{Caneva:1980}
Caneva, K.~L. (1980).
\newblock Amp\`ere, the etherians, and the oersted connexion.
\newblock {\em The British Journal for the History of Science\/}, {\em
  13\/}(2), 121--138.

\bibitem[{Chalmers(1999)}]{Chalmers:1999}
Chalmers, A.~F. (1999).
\newblock {\em What Is This Thing Called Science?\/}.
\newblock Hackett Pub Co Inc, 3 ed.

\bibitem[{Chi(2009)}]{Chi:2009}
Chi, M. T.~H. (2009).
\newblock Active-constructive-interactive: A conceptual framework for
  differentiating learning activities.
\newblock {\em Topics in Cognitive Science\/}, {\em 1\/}(1), 73--105.

\bibitem[{Chi et~al.(1989)Chi, Bassok, Lewis, Reimann, \&
  Glaser}]{ChiEtAll:1989}
Chi, M. T.~H., Bassok, M., Lewis, M.~W., Reimann, P., \& Glaser, R. (1989).
\newblock Self-explanations: How students study and use examples in learning to
  solve problems.
\newblock {\em Cognitive Science\/}, {\em 13\/}(2), 145 -- 182.

\bibitem[{Chin \& Osborne(2010{\natexlab{a}})}]{Chin:2010b}
Chin, C., \& Osborne, J. (2010{\natexlab{a}}).
\newblock Students' questions and discursive interaction: Their impact on
  argumentation during collaborative group discussions in science.
\newblock {\em J. Res. Sci. Teach.\/}, {\em 47\/}(7), 883--908.

\bibitem[{Chin \& Osborne(2010{\natexlab{b}})}]{Chin:2010a}
Chin, C., \& Osborne, J. (2010{\natexlab{b}}).
\newblock Supporting argumentation through students' questions: Case studies in
  science classrooms.
\newblock {\em J. Learn. Sci.\/}, {\em 19\/}(2), 230--284.

\bibitem[{Coletta et~al.(2008)Coletta, Phillips, Savinainen, \&
  Steinert}]{Coletta:2008}
Coletta, V.~P., Phillips, J.~A., Savinainen, A., \& Steinert, J.~J. (2008).
\newblock Comment on \char39{}the effects of students\char39{} reasoning
  abilities on conceptual understandings and problem-solving skills in
  introductory mechanics\char39{}.
\newblock {\em Eur. J. Phys.\/}, {\em 29\/}, L25--L27.

\bibitem[{Davies(1996)}]{Davies:1996}
Davies, P. C.~W. (1996).
\newblock Einstein's greatest mistake?
\newblock {\em Astrophys. Space Sci.\/}, {\em 244\/}, 219--227.

\bibitem[{diSessa(2008)}]{diSessa:2008}
diSessa, A.~A. (2008).
\newblock A ``theory bite'' on the meaning of scientific inquiry: A companion
  to kuhn and pease.
\newblock {\em Cognition Instruct.\/}, {\em 26\/}(4), 560--566.

\bibitem[{Docktor \& Mestre(2011)}]{DocktorAndMestre:xxxx}
Docktor, J.~L., \& Mestre, J.~P. (2011).
\newblock A synthesis of discipline-based education research in physics.
\newblock
  \url{http://www7.nationalacademies.org/bose/DBER_Docktor_October_Paper.pdf}.

\bibitem[{Dyson(1991)}]{Dyson:1991}
Dyson, F.~J. (1991).
\newblock ``to teach or not to teach,'' {F}reeman {J}. {D}yson's acceptance
  speech for the 1991 {O}ersted {M}edal presented by the {A}merican
  {A}ssociation of {P}hysics {T}eachers, 22 january 1991.
\newblock {\em Am. J. Phys.\/}, {\em 59\/}, 491--495.

\bibitem[{Feldman(2010)}]{Feldman:2010}
Feldman, B.~J. (2010).
\newblock The collapse of the i-35w bridge in minneapolis.
\newblock {\em Phys. Teach.\/}, {\em 48\/}(8), 541--542.

\bibitem[{Finkelstein et~al.(2009)Finkelstein, Mazur, \& Lasry}]{Lasry:2009b}
Finkelstein, N., Mazur, E., \& Lasry, N. (2009).
\newblock What should we expect students to learn?
\newblock {\em Phys. Teach.\/}, {\em 47\/}(8), 484--484.

\bibitem[{Fortus(2009)}]{Fortus:2009}
Fortus, D. (2009).
\newblock The importance of learning to make assumptions.
\newblock {\em Sci. Educ.\/}, {\em 93\/}(1), 86--108.

\bibitem[{Gaigher et~al.(2007)Gaigher, Rogan, \& Braun}]{GaigherEtAl:2007}
Gaigher, E., Rogan, J., \& Braun, M. (2007).
\newblock Exploring the development of conceptual understanding through
  structured problem-solving in physics.
\newblock {\em Int. J. Sci. Educ.\/}, {\em 29\/}, 1089--1110.

\bibitem[{Giunta(2001)}]{Giunta:2001}
Giunta, C.~J. (2001).
\newblock Using history to teach scientific method: The role of errors.
\newblock {\em J. Chem. Educ.\/}, {\em 78\/}(5), 623.

\bibitem[{G\l{}azek \& Grayson(2008)}]{Glazek:2008}
G\l{}azek, S., \& Grayson, D. (2008).
\newblock Fine points on productive learning.
\newblock {\em Phys. Today\/}, {\em 61\/}(11), 11--12.

\bibitem[{Graesser \& Person(1994)}]{GraesserPerson:1994}
Graesser, A.~C., \& Person, N.~K. (1994).
\newblock Question asking during tutoring.
\newblock {\em American Educational Research Journal\/}, {\em 31\/}(1),
  104--137.

\bibitem[{Hamed(2008)}]{Hamed:2008}
Hamed, K.~M. (2008).
\newblock Do you prefer to have the text or a sheet with your physics exams?
\newblock {\em Phys. Teach.\/}, {\em 46\/}, 290--293.

\bibitem[{Hammer et~al.(2011)Hammer, Gupta, \& Redish}]{HammerEtAl:2011}
Hammer, D., Gupta, A., \& Redish, E.~F. (2011).
\newblock On static and dynamic intuitive ontologies.
\newblock {\em Journal of the Learning Sciences\/}, {\em 20\/}(1), 163--168.

\bibitem[{Harper et~al.(2003)Harper, Etkina, \& Lin}]{HarperEtAll:2003}
Harper, K.~A., Etkina, E., \& Lin, Y. (2003).
\newblock Encouraging and analyzing student questions in a large physics
  course: Meaningful patterns for instructors.
\newblock {\em J. Res. Sci. Teach.\/}, {\em 40\/}, 776--791.

\bibitem[{Hatano \& Inagaki(1986)}]{HatanoInagaki:1986}
Hatano, G., \& Inagaki, K. (1986).
\newblock Two courses of expertise.
\newblock In H.~Stevenson, H.~Azuma, \& K.~Hakuta (Eds.) {\em Child development
  and education in japan\/}, (pp. 262--272). W. H. Freeman and Company, New
  York. Available at \url{http://hdl.handle.net/2115/25206}.

\bibitem[{Hawkins(2011)}]{Hawkins:2011}
Hawkins, J. (2011).
\newblock In preparation.

\bibitem[{Heller \& Reif(1984)}]{HellerReif:1984}
Heller, J.~I., \& Reif, F. (1984).
\newblock Prescribing effective human problem-solving processes: Problem
  description in physics.
\newblock {\em Cognition Instruct.\/}, {\em 1\/}(2), 177--216.

\bibitem[{Henderson \& Dancy(2009)}]{HendersonDancy:2009}
Henderson, C., \& Dancy, M.~H. (2009).
\newblock Impact of physics education research on the teaching of introductory
  quantitative physics in the united states.
\newblock {\em Phys. Rev. ST Phys. Educ. Res.\/}, {\em 5\/}, 020107:1--9.

\bibitem[{Heron \& Meltzer(2005)}]{HeronMeltzer:2005}
Heron, P. R.~L., \& Meltzer, D.~E. (2005).
\newblock The future of physics education research: intellectual challenges and
  practical concerns.
\newblock {\em Am. J. Phys.\/}, {\em 73\/}, 390--394.

\bibitem[{Hewitt(2011)}]{Hewitt:2011}
Hewitt, P.~G. (2011).
\newblock Equations as guides to thinking and problem solving.
\newblock {\em Phys. Teach.\/}, {\em 49\/}(5), 264--264.

\bibitem[{Higgins(1997)}]{Higgins:1997}
Higgins, K.~M. (1997).
\newblock The effect of year-long instruction in mathematical problem solving
  on middle-school students' attitudes, beliefs, and abilities.
\newblock {\em The Journal of Experimental Education\/}, {\em 66\/}(1), 5--28.

\bibitem[{Hobson(2009)}]{Hobson:2009}
Hobson, A. (2009).
\newblock Response to ``the scandal of quantum mechanics,'' by {N}. {G}. van
  kampen [am. j. phys. \textbf{76}, 989--990 (2008)].
\newblock {\em Am. J. Phys.\/}, {\em 77\/}(4), 293--293.

\bibitem[{Hoellwarth \& Moelter(2011)}]{HoellwarthMoelter:2011}
Hoellwarth, C., \& Moelter, M.~J. (2011).
\newblock The implications of a robust curriculum in introductory mechanics.
\newblock {\em Am. J. Phys.\/}, {\em 79\/}(5), 540--545.

\bibitem[{Hoellwarth et~al.(2005)Hoellwarth, Moelter, \&
  Knight}]{HoellwarthEtAll:2005}
Hoellwarth, C., Moelter, M.~J., \& Knight, R.~D. (2005).
\newblock A direct comparison of conceptual learning and problem solving
  ability in traditional and studio style classrooms.
\newblock {\em Am. J. Phys.\/}, {\em 73\/}, 459--462.

\bibitem[{Hrepic et~al.(2007)Hrepic, Zollman, \& Rebello}]{HrepicEtAll:2007}
Hrepic, Z., Zollman, D., \& Rebello, N. (2007).
\newblock Comparing students' and experts' understanding of the content of a
  lecture.
\newblock {\em Journal of Science Education and Technology\/}, {\em 16\/},
  213--224.

\bibitem[{Hutzler et~al.(2004)Hutzler, Delaney, Weaire, \&
  MacLeod}]{HutzlerEtAl:2004}
Hutzler, S., Delaney, G., Weaire, D., \& MacLeod, F. (2004).
\newblock Rocking newton's cradle.
\newblock {\em Am. J. Phys.\/}, {\em 72\/}, 1508--1516.

\bibitem[{Jackson(1999)}]{Jackson:1999}
Jackson, J.~D. (1999).
\newblock Maxwell's displacement current revisited.
\newblock {\em Eur. J. Phys.\/}, {\em 20\/}(6), 495.

\bibitem[{Jackson(2000)}]{JacksonReplyRoche:2000}
Jackson, J.~D. (2000).
\newblock Reply to comment by j roche on `maxwell's displacement current
  revisited'.
\newblock {\em Eur. J. Phys.\/}, {\em 21\/}(4), L29.

\bibitem[{Janssen(2002)}]{Janssen:2002}
Janssen, M. (2002).
\newblock Reconsidering a scientific revolution: The case of einstein versus
  lorentz.
\newblock {\em Physics in Perspective (PIP)\/}, {\em 4\/}, 421--446.

\bibitem[{Kahneman(2002)}]{Kahneman:2002}
Kahneman, D. (2002).
\newblock Maps of bounded rationality: A perspective on intuitive judgment and
  choice.
\newblock {\em Economic Sciences Nobel Price Lecture, December 8, 2002 (pages
  449-489). 
The lecture is available at
  {\scriptsize \url{http://nobelprize.org/nobel_prizes/economics/laureates/2002/kahnemann-l%ecture.pdf}\/}}.

\bibitem[{Kamal(2011)}]{Kamal:2011}
Kamal, A.~A. (2011).
\newblock {\em 1000 Solved Problems in Classical Physics\/}.
\newblock Springer Berlin Heidelberg.

\bibitem[{Kim \& Pak(2002)}]{KimPak:2002}
Kim, E., \& Pak, S. (2002).
\newblock Students do not overcome conceptual difficulties after solving 1000
  tra ditional problems.
\newblock {\em Am. J. Phys.\/}, {\em 70\/}(7), 759--765.

\bibitem[{Kipnis(2011)}]{Kipnis:2010}
Kipnis, N. (2011).
\newblock Errors in science and their treatment in teaching science.
\newblock {\em Sci. \& Educ.\/}, {\em 20\/}, 655--685.

\bibitem[{Klein(2007{\natexlab{a}})}]{KleinDebate3:2007}
Klein, D. (2007{\natexlab{a}}).
\newblock Reply to comments on ``{S}chool math books, nonsense, and the
  {N}ational {S}cience {F}oundation,'' by {D}avid {K}lein\cite{Klein:2007}.
\newblock {\em Am. J. Phys.\/}, {\em 75\/}, 776--778.

\bibitem[{Klein(2007{\natexlab{b}})}]{Klein:2007}
Klein, D. (2007{\natexlab{b}}).
\newblock School math books, nonsense, and the national science foundation.
\newblock {\em Am. J. Phys.\/}, {\em 75\/}, 101--102.

\bibitem[{Knight(2008)}]{Knight:2008}
Knight, R.~D. (2008).
\newblock {\em Instructor Guide. Physics for scientists and engineers. A
  strategic approach, second edition\/}.
\newblock Pearson, Addison-Wesley.

\bibitem[{Kolodner(1992)}]{Kolodner:1992}
Kolodner, J.~L. (1992).
\newblock An introduction to case-based reasoning.
\newblock {\em Artif. Intell. Review\/}, {\em 6\/}, 3--34.

\bibitem[{Kost-Smith et~al.(2010)Kost-Smith, Pollock, \&
  Finkelstein}]{KostSmithEtAl:2010}
Kost-Smith, L.~E., Pollock, S.~J., \& Finkelstein, N.~D. (2010).
\newblock Gender disparities in second-semester college physics: The
  incremental effects of a ``smog of bias''.
\newblock {\em Phys. Rev. ST Phys. Educ. Res.\/}, {\em 6\/}, 020112:1--17.

\bibitem[{Kuhn et~al.(2008)Kuhn, Iordanou, Pease, \& Wirkala}]{Kuhn:2008}
Kuhn, D., Iordanou, K., Pease, M., \& Wirkala, C. (2008).
\newblock Beyond control of variables: What needs to develop to achieve skilled
  s cientific thinking?
\newblock {\em Cognitive Development\/}, {\em 23\/}(4), 435 -- 451.

\bibitem[{Kuhn et~al.(2009)Kuhn, Pease, \& Wirkala}]{Kuhn:2009}
Kuhn, D., Pease, M., \& Wirkala, C. (2009).
\newblock Coordinating the effects of multiple variables: A skill fundamental
  to scientific thinking.
\newblock {\em Journal of Experimental Child Psychology\/}, {\em 103\/}(3), 268
  -- 284.

\bibitem[{Labudde et~al.(1988)Labudde, Reif, \& Quinn}]{LabuddeReifQuinn:1988}
Labudde, P., Reif, F., \& Quinn, L. (1988).
\newblock Facilitation of scientific concept learning by interpretation
  procedures and diagnosis.
\newblock {\em Int. J. Sci. Educ.\/}, {\em 66\/}(2), 181--221.

\bibitem[{Lalo\"{e}(2001)}]{Lalo:2001}
Lalo\"{e}, F. (2001).
\newblock Do we really understand quantum mechanics? strange correlations,
  paradoxes, and theorems.
\newblock {\em Am. J. Phys.\/}, {\em 69\/}(6), 655--701.

\bibitem[{Lange(2011)}]{Lange:2011}
Lange, M. (2011).
\newblock Why do forces add vectorially? a forgotten controversy in the
  foundations of classical mechanics.
\newblock {\em Am. J. Phys.\/}, {\em 79\/}(4), 380--388.

\bibitem[{Larkin et~al.(1980)Larkin, McDermott, Simon, \&
  Simon}]{LarkinEtAl:1980}
Larkin, J., McDermott, J., Simon, D.~P., \& Simon, H.~A. (1980).
\newblock Expert and novice performance in solving physics problems.
\newblock {\em Science\/}, {\em 208\/}(4450), 1335--1342.

\bibitem[{Lasry et~al.(2009)Lasry, Finkelstein, \& Mazur}]{Lasry:2009a}
Lasry, N., Finkelstein, N., \& Mazur, E. (2009).
\newblock Are most people too dumb for physics?
\newblock {\em Phys. Teach.\/}, {\em 47\/}(7), 418--422.

\bibitem[{Lindley(2001)}]{Lindley:2001}
Lindley, D. (2001).
\newblock {\em Boltzmanns Atom: The Great Debate That Launched A Revolution In
  Physics\/}.
\newblock Free Press.

\bibitem[{Litzinger et~al.(2011)Litzinger, Lattuca, Hadgraft, \&
  Newstetter}]{LitzingerEtAl:2011}
Litzinger, T.~A., Lattuca, L.~R., Hadgraft, R.~G., \& Newstetter, W.~C. (2011).
\newblock Engineering education and the development of expertise.
\newblock {\em J. Eng. Educ.\/}, {\em 100\/}, 123--150.

\bibitem[{Marshall \& Carrejo(2008)}]{MarshallCarrejo:2008}
Marshall, J.~A., \& Carrejo, D.~J. (2008).
\newblock Students' mathematical modeling of motion.
\newblock {\em J. Res. Sci. Teach.\/}, {\em 45\/}(2), 153--173.

\bibitem[{Mason \& Singh(2010)}]{MasonSingh:2010}
Mason, A., \& Singh, C. (2010).
\newblock Helping students learn effective problem solving strategies by
  reflecting with peers.
\newblock {\em Am. J. Phys.\/}, {\em 78\/}, 748--754.

\bibitem[{McDermott \& Shaffer(1992)}]{McDermott:1992:p1}
McDermott, L.~C., \& Shaffer, P.~S. (1992).
\newblock Research as a guide for curriculum development: An example from
  introductory electricity. part i: Investigation of student understanding.
\newblock {\em Am. J. Phys.\/}, {\em 60\/}(11), 994--1003.

\bibitem[{Meredith \& Marrongelle(2008)}]{Meredith:2008}
Meredith, D.~C., \& Marrongelle, K.~A. (2008).
\newblock How students use mathematical resources in an electrostatics context.
\newblock {\em Am. J. Phys.\/}, {\em 76\/}, 570--578.

\bibitem[{Millar(2007)}]{KleinDebate2:2007}
Millar, T. (2007).
\newblock Comment on ``{S}chool math books, nonsense, and the {N}ational
  {S}cience {F}oundation,'' by {D}avid {K}lein\cite{Klein:2007}.
\newblock {\em Am. J. Phys.\/}, {\em 75\/}, 775--776.

\bibitem[{Mualem \& Eylon(2007)}]{MualemEylon:2007}
Mualem, R., \& Eylon, B.~S. (2007).
\newblock 'physics with a smile'-explaining phenomena with a qualitative
  problem-solving strategy.
\newblock {\em Phys. Teach.\/}, {\em 45\/}, 158--163.
\newblock See also references there in.

\bibitem[{Nickerson(1988)}]{Nickerson:1988}
Nickerson, R.~S. (1988).
\newblock On improving thinking through instruction.
\newblock {\em Review of Research in Education\/}, {\em 15\/}, 3--57.

\bibitem[{Ogilvie(2009)}]{Ogilvie:2009}
Ogilvie, C.~A. (2009).
\newblock Changes in students' problem-solving strategies in a course that
  includes context-rich, multifaceted problems.
\newblock {\em Phys. Rev. ST Phys. Educ. Res.\/}, {\em 5\/}, 020102:1--14.

\bibitem[{Olness \& Scalise(2011)}]{OlnessAndScalise:2011}
Olness, F., \& Scalise, R. (2011).
\newblock Regularization, renormalization, and dimensional analysis:
  Dimensional regularization meets freshman e\&m.
\newblock {\em Am. J. Phys.\/}, {\em 79\/}, 306--312.

\bibitem[{Osborne(2010)}]{Osborne:2010}
Osborne, J. (2010).
\newblock Arguing to learn in science: The role of collaborative, critical
  discourse.
\newblock {\em Science\/}, {\em 328\/}(5977), 463--466.

\bibitem[{Peierls(1979)}]{Peierls:1979}
Peierls, R. (1979).
\newblock {\em Surprises in Theoretical Physics\/}.
\newblock Princeton University Press.

\bibitem[{Peierls(1991)}]{Peierls:1991}
Peierls, R. (1991).
\newblock {\em More Surprises in Theoretical Physics\/}.
\newblock Princeton University Press.

\bibitem[{Plotnitsky(2010)}]{Plotnitsky:2010}
Plotnitsky, A. (2010).
\newblock {\em Epistemology and Probability\/}.
\newblock Springer New York.

\bibitem[{Polya(1973)}]{Polya:1945}
Polya, G. (1973).
\newblock {\em How to Solve it. A new aspect of mathematical method\/}.
\newblock Princeton University Press, 2nd. ed.

\bibitem[{Reif(2008)}]{Reif:2008}
Reif, F. (2008).
\newblock {\em Applying cognitive science to education : thinking and learning
  in scientific and other complex domains\/}.
\newblock MIT Press.

\bibitem[{Reif \& Allen(1992)}]{ReifAllen:1992}
Reif, F., \& Allen, S. (1992).
\newblock Cognition for interpreting scientific concepts: A study of
  acceleration.
\newblock {\em Cognition Instruct.\/}, {\em 9\/}(1), 1--44.

\bibitem[{Reif \& Heller(1982)}]{ReifHeller:1982}
Reif, F., \& Heller, J.~I. (1982).
\newblock Knowledge structure and problem solving in physics.
\newblock {\em Educ. Psychol.\/}, {\em 17\/}(2), 102--127.

\bibitem[{Roche(2000)}]{RocheReplyJackson:2000}
Roche, J. (2000).
\newblock Reply to j d jackson's `maxwell's displacement current revisited'.
\newblock {\em Eur. J. Phys.\/}, {\em 21\/}(4), L27.

\bibitem[{Rojas()}]{Rojas:arxiv:2010}
Rojas, S.~. (????).
\newblock A non-standard approach to introduce simple harmonic motion.
\newblock eprint arXiv:1011.0687,\\ {\url{http://arxiv.org/abs/1011.0687v1} }.

\bibitem[{Rojas(2008)}]{Rojas:2008}
Rojas, S. (2008).
\newblock On the need to enhance physical insight via mathematical reasoning.
\newblock {\em Rev. Mex. F{\'{\i}}s. E\/}, {\em 54\/}, 75--80.

\bibitem[{Rojas(2010)}]{Rojas:2010rmf}
Rojas, S. (2010).
\newblock On the teaching and learning of physics problem solving.
\newblock {\em Rev. Mex. F{\'{\i}}s. E\/}, {\em 56\/}, 22--28.

\bibitem[{Sabella \& Redish(2007)}]{SabellaRedish:2007}
Sabella, M.~S., \& Redish, E.~F. (2007).
\newblock Knowledge organization and activation in physics problem solving.
\newblock {\em Am. J. Phys.\/}, {\em 75\/}, 1017--1029.

\bibitem[{Saccomandi(2010)}]{Saccomandi:2010}
Saccomandi, G. (2010).
\newblock On the motion of the centre of mass of a system of particles.
\newblock {\em Eur. J. Phys.\/}, {\em 31\/}, 657--670.

\bibitem[{Sandin(1973)}]{Sandin:1973}
Sandin, T.~R. (1973).
\newblock Nonconservation of linear momentum in ballistic pendulums.
\newblock {\em Am. J. Phys.\/}, {\em 41\/}, 426--427.

\bibitem[{Schoenfeld(1980)}]{Schoenfeld:1980}
Schoenfeld, A.~H. (1980).
\newblock Teaching problem-solving skills.
\newblock {\em The American Mathematical Monthly\/}, {\em 87\/}, 794--805.

\bibitem[{Schoenfeld(1992)}]{Schoenfeld:1992}
Schoenfeld, A.~H. (1992).
\newblock Learning to think mathematically: Problem solving, metacognition, and
  sense-making in mathematics.
\newblock In {\em {D. Grouws (Ed.), Handbook for Research on Mathematics
  Teaching and Learning }\/}, (pp. 334--370). MacMillan.

\bibitem[{Schwartz et~al.(2005{\natexlab{a}})Schwartz, Bransford, \&
  Sears}]{SchwartzEtAll:2005}
Schwartz, D.~L., Bransford, J.~D., \& Sears, D. (2005{\natexlab{a}}).
\newblock Efficiency and innovation in transfer.
\newblock In J.~P. Mestre (Ed.) {\em Transfer of learning from a modern
  multidisciplinary perspective\/}, (pp. 1--52). Information Age Publishing.

\bibitem[{Schwartz et~al.(2005{\natexlab{b}})Schwartz, Taylor, \&
  Jay}]{SchwartzEtAll:MathPhys:2005}
Schwartz, D.~L., Taylor, M., \& Jay, P. (2005{\natexlab{b}}).
\newblock How mathematics propels the development of physical knowledge.
\newblock {\em J. Cogn. Dev.\/}, {\em 6\/}, 65--88.

\bibitem[{Shaffer \& McDermott(1992)}]{McDermott:1992:p2}
Shaffer, P.~S., \& McDermott, L.~C. (1992).
\newblock Research as a guide for curriculum development: An example from
  introductory electricity. part ii: Design of instructional strategies.
\newblock {\em Am. J. Phys.\/}, {\em 60\/}(11), 1003--1013.

\bibitem[{Shaffer \& McDermott(2005)}]{Shaffer:2005}
Shaffer, P.~S., \& McDermott, L.~C. (2005).
\newblock A research-based approach to improving student understanding of the
  vector nature of kinematical concepts.
\newblock {\em Am. J. Phys.\/}, {\em 73\/}(10), 921--931.

\bibitem[{{Sherin}(2001)}]{Sherin:2001}
{Sherin}, B.~L. (2001).
\newblock {How Students Understand Physics Equations}.
\newblock {\em Cognition Instruct.\/}, {\em 19\/}, 479--541.

\bibitem[{Singh(2002)}]{Singh:2002}
Singh, C. (2002).
\newblock When physical intuition fails.
\newblock {\em Am. J. Phys.\/}, {\em 70\/}, 1103--1109.

\bibitem[{Slotta(2011)}]{Slotta:2011}
Slotta, J.~D. (2011).
\newblock In defense of chi's ontological incompatibility hypothesis.
\newblock {\em Journal of the Learning Sciences\/}, {\em 20\/}(1), 151--162.

\bibitem[{Sobel(2009{\natexlab{a}})}]{Sobel:2009a}
Sobel, M. (2009{\natexlab{a}}).
\newblock Physics for the non-scientist: A middle way.
\newblock {\em Phys. Teach.\/}, {\em 47\/}(6), 346--349.

\bibitem[{Sobel(2009{\natexlab{b}})}]{Sobel:2009b}
Sobel, M. (2009{\natexlab{b}}).
\newblock Response to ``are most people too dumb for physics?''.
\newblock {\em Phys. Teach.\/}, {\em 47\/}(7), 422--423.

\bibitem[{Strogatz et~al.(2005)Strogatz, Abrams, McRobie, Eckhardt, \&
  Ott}]{StrogatzEtAl:2005}
Strogatz, S.~H., Abrams, D.~M., McRobie, A., Eckhardt, B., \& Ott, E. (2005).
\newblock Crowd synchrony on the millennium bridge.
\newblock {\em Nature\/}, {\em 438\/}, 43--44.

\bibitem[{Swendsen(2011)}]{Swendsen:2011}
Swendsen, R.~H. (2011).
\newblock How physicists disagree on the meaning of entropy.
\newblock {\em Am. J. Phys.\/}, {\em 79\/}(4), 342--348.

\bibitem[{Taber(2009)}]{Taber:2009}
Taber, K.~S. (2009).
\newblock Maths should be the last thing we teach.
\newblock {\em Phys. Educ.\/}, {\em 44\/}(4), 336.

\bibitem[{Thompson et~al.(2009)Thompson, Christensen, Pollock, Bucy, \&
  Mountcastle}]{Thompson:2009}
Thompson, J.~R., Christensen, W.~M., Pollock, E.~B., Bucy, B.~R., \&
  Mountcastle, D.~B. (2009).
\newblock Student understanding of thermal physics concepts and the underlying
  mathematics in the upper division.
\newblock In {\em {Proceedings of the Frontiers in Science Education Research
  Conference, 24-24 March 2009. Eastern Mediterranean University, Famagusta,
  North Cyprus}\/}.

\bibitem[{van Kampen(2008)}]{Kampen:2008}
van Kampen, N.~G. (2008).
\newblock The scandal of quantum mechanics.
\newblock {\em Am. J. Phys.\/}, {\em 76\/}(11), 989--990.

\bibitem[{van Weeren et~al.(1982)van Weeren, de~Mul, Peters, Kramers-Pals, \&
  Roossink}]{Weeren:1982}
van Weeren, J. H.~P., de~Mul, F. F.~M., Peters, M.~J., Kramers-Pals, H., \&
  Roossink, H. (1982).
\newblock Teaching problem-solving in physics: A course in electromagnetism.
\newblock {\em Am. J. Phys.\/}, {\em 50\/}, 725--732.

\bibitem[{Wagner(2006)}]{Wagner:2006}
Wagner, J.~F. (2006).
\newblock Transfer in pieces.
\newblock {\em Cognition Instruct.\/}, {\em 24\/}, 1 -- 71.

\bibitem[{Wallace \& Chasteen(2010)}]{WallaceChasteen:2010}
Wallace, C.~S., \& Chasteen, S.~V. (2010).
\newblock Upper-division students' difficulties with amp\`ere's law.
\newblock {\em Phys. Rev. ST Phys. Educ. Res.\/}, {\em 6\/}, 020115:1--8.

\bibitem[{Walsh et~al.(2007)Walsh, Howard, \& Bowe}]{Walsh:2007}
Walsh, L.~N., Howard, R.~G., \& Bowe, B. (2007).
\newblock Phenomenographic study of students\char39{} problem solving
  approaches in physics.
\newblock {\em Phys. Rev. ST Phys. Educ. Res.\/}, {\em 3\/}, 020108:1--12.

\bibitem[{Weinberg(2005)}]{Weinberg:2005}
Weinberg, S. (2005).
\newblock Einstein's mistakes.
\newblock {\em Phys. Today\/}, {\em 58\/}(11), 31--35.

\bibitem[{Whittaker(1910)}]{Whittaker:1910}
Whittaker, E.~T. (1910).
\newblock {\em A history of the theories of aether and electricity : from the
  age of Descartes to the close of the nineteenth century\/}.
\newblock Longmans, Green and CO.. Available at
  \url{http://www.archive.org/details/historyoftheorie00whitrich}.

\bibitem[{Wieman(2007)}]{Wieman:2007a}
Wieman, C. (2007).
\newblock The back page: The ``curse of knowledge'', or why intuition about
  teaching often fails.
\newblock {\em APS News\/}, {\em 16\/}(10), 8--8.

\bibitem[{Zee \& Minstrell(1997)}]{ZeeMinstrell:1997}
Zee, E.~V., \& Minstrell, J. (1997).
\newblock Using questioning to guide student thinking.
\newblock {\em The Journal of the Learning Sciences\/}, {\em 6\/}(2), 227--269.

\end{thebibliography}
\end{document}